\documentclass[a4paper,onecolumn,showpacs,superscriptaddress,groupedaddress]{revtex4}
\usepackage{ae}
\usepackage[T1]{fontenc}
\usepackage[ansinew]{inputenc}
\usepackage{amsmath}
\usepackage{amssymb}
\usepackage{graphicx}
\usepackage[caption=false]{subfig}
\usepackage{color}
\usepackage[colorlinks]{hyperref}
\usepackage{lscape}
\usepackage{natbib}
\begin{document}
\title{Decoherence Dynamics of Measurement-Induced Nonlocality and comparison with Geometric Discord for two qubit systems}

\author{Ajoy Sen}
\email{ajoy.sn@gmail.com}
\affiliation{Department of Applied Mathematics, University of Calcutta, 92, A.P.C. Road, Kolkata-700009, India}
\author{Debasis Sarkar}
\email{dsappmath@caluniv.ac.in}
\affiliation{Department of Applied Mathematics, University of Calcutta, 92, A.P.C. Road, Kolkata-700009, India}
\author{Amit Bhar}
\email{bhar.amit@yahoo.com}
\affiliation{Department of Mathematics, Jogesh Chandra Chaudhuri College, 30, Prince Anwar Shah Road, Kolkata-700033, India}

\begin{abstract}
We check the decoherence dynamics of Measurement-induced Nonlocality(in short, MIN) and compare it with geometric discord for two qubit systems. There are quantum states, on which the action of dephasing channel cannot destroy MIN in finite or infinite time. We check the additive dynamics of MIN on a qubit state under two independent noise. Geometric discord also follows such additive dynamics like quantum discord. We have further compared non-Markovian evolution of MIN and geometric discord under dephasing and amplitude damping noise for pure state and it shows distinct differences between their dynamics.
\end{abstract}
\date{\today}
\pacs{03.67. -a, 03.65.Ud.}
\maketitle

\section{Introduction}

Separability, Nonlocality and quantumness these are the three most discussed areas in Quantum Information theory. While the first one has a definite identifier in terms of entanglement measures; the last two lacks definite identifiers as they can exist in various forms. But there is one thing in common among all of them. All help us to understand the state space structures of quantum systems from a very different perspectives. The idea behind discord \cite{zurek,dakic} was to differentiate classicality from quantumness. There are separable states which have non-zero value of discord, indicating existence of quantumness beyond entanglement.  Measurement-induced nonlocality(MIN) as introduced by Luo and Fu \cite{luo} is another type of correlation which reveals a type of nonlocality that also exists in composite quantum systems. There are classical states(i.e., states with vanishing discord) with non-zero measurement induced nonlocality \cite{luo,guo}. Since our classical world is local so existence of non-zero MIN for classical states exhibits a kind of non-classical trait in those states.

Study on the behavior of non-classical correlations under decoherence  started with the strange observation of Entanglement sudden death(in short, ESD) \cite{eberly}. For some class of initial states ESD is observed under independent Markovian decoherence but similar behavior is not observed in case of quantum discord \cite{werlang}. In fact quantum discord decays monotonically under such decoherences and sudden death of discord can not occur under Markovian decoherence \cite{ferraro}. Yu and Eberly had also pointed out non-additivity of decoherence channel under entanglement dynamics but opposite result is observed in case of quantum discord. The dynamics of quantum discord under decoherence channels and Pauli channels is also well studied in \cite{maziero,werlang}. Geometric discord(in short, GD), as introduced by Dakic and Vedral \cite{dakic} is a way to visualize discord  more geometrically. Behavior of different measures of discord have quite different characters \cite{bellomo,2}. The value of quantum discord for pure bipartite states matches with the entanglement of the state, whereas, the value of Geometric discord matches with the value of MIN. Here we will mainly investigate the decoherence dynamics of MIN under some dissipative(depolarizing, amplitude damping) and non-dissipative(dephasing) channels for two qubit systems. They are some of the most important general unital and non-unital qubit channels. We take particular initial conditions and compare the dynamics with the corresponding dynamics of geometric discord. We will investigate the additivity of decoherence channels under MIN and GD dynamics. We will further compare the non-Markovian evolution of MIN and geometric discord under dephasing and amplitude damping noise for pure state and obtain some distinguishable differences with the earlier results. In our study, we will use original version of MIN as introduced by Luo and Fu. However there are entropic formulation of MIN. Few results regarding this area including studies in dynamics of discord in various approaches can be found in \cite{Guo-Feng,Ming,1,3,4,5,6}.

\section{Preliminary notions on MIN and Geometric Discord}
Let $\rho_{AB}$ be any bipartite state shared between two parties A and B. Then MIN (denoted by $N(\rho_{AB})$) is defined as \cite{luo},
\begin{equation}
N(\rho_{AB}):=\max_{\Pi^{\text{A}}}\parallel \rho_{AB}-\Pi^{\text{A}}\otimes I^B(\rho_{AB})\parallel^2
\end{equation}
where maximum is over all von Neumann measurements $\Pi^{\text{A}} = \{ \Pi^A_k \}$ on party A which do not disturb $\rho_{\text{A}}$, the local density matrix  of A, i.e., $\Sigma_{k}\Pi_{k}^{\text{A}}\rho_{\text{A}}\Pi_{k}^{\text{A}}=\rho_{\text{A}}$ and $\|.\|$ is taken as Hilbert Schmidt norm (i.e. $\parallel X \parallel=[\text{tr}(X^{\dag}X)]^{\frac{1}{2}}$), $I^B$ is the identity operator acting on party B.
On the other hand Geometric discord for a quantum state $\rho_{AB}$ is defined as \cite{dakic},
\begin{equation}\label{gddef}
D_{g}(\rho_{AB})=\min_{\chi_{AB}\in\Omega_{0}}\parallel \rho_{AB}-\chi_{AB}\parallel^2
\end{equation}
where $\Omega_{0}$ is the set of all zero discord states and $\parallel.\parallel$ is also the usual Hilbert-Schmidt norm.
Any state $\rho_{\text{AB}}$ of a two qubit quantum system can be written in the form,
\begin{equation}
\begin{split}\label{qubitstate}
\rho_{\text{AB}}=\frac{1}{\sqrt{4}}\frac{I^{\text{A}}}{\sqrt{2}}\otimes{\frac{I^{\text{B}}}{\sqrt{2}}}+\sum_{i=1}^{3}x_{i}X_{i}^{\text{A}}\otimes{\frac{I^{\text{B}}}{\sqrt{2}}}
+\frac{I^{\text{A}}}{\sqrt{2}}\otimes{\sum_{i=1}^{3}y_{i}Y_{i}^{\text{B}}}+\sum_{i=1}^{3}\sum_{j=1}^{3}t_{ij}X_{i}^{\text{A}}\otimes{Y_{j}^{\text{B}}}
\end{split}
\end{equation}
where $\{X_{i}^{\text{A}}:\,$ i=0,1,2,3\} and $\{Y_{j}^{\text{B}}:\,{ j=0,1,2,3}\}$ are the orthonormal Hermitian operator bases for $L(H^{\text{A}})$ and $L(H^{\text{B}})$ respectively, with $X_{0}^{\text{A}}=Y_{0}^{\text{B}}=I^{\text{A}}/\sqrt{2}$. Here we will use the Pauli matrices (with normalization factor $\frac{1}{\sqrt{2}}$) as the operator base, i.e., $X_i^{\text{A}}=Y_i^{\text{B}}=\frac{\sigma_i}{\sqrt{2}}$. For this state (\ref{qubitstate}) MIN \cite{luo,mirafzali} and geometric discord \cite{dakic} can be explicitly written as,
\begin{equation}\label{min}
N(\rho_{^{\text{AB}}})=
\begin{cases}
\text{tr}TT^{t}-\frac{1}{\|\textbf{x}\|^{2}}\textbf{x}^{t}TT^{t}\textbf{x}&  \text{if}~\textbf{x}\neq{\textbf{0}}\\
\text{tr}TT^{t}-\lambda_{min} & \text{if}~\textbf{x}=\textbf{0}\\
\end{cases}
\end{equation}
\begin{equation}\label{discord}
 D_{g}(\rho_{^{\text{AB}}})=(\parallel \overrightarrow{x}\parallel^{2}+\parallel T\parallel^{2}-\lambda_{max})
\end{equation}
where the matrix $T=(t_{ij})_{3\times 3}$ is the correlation matrix with  $\lambda_{min}$ being minimum eigenvalue of $TT^{t}$ and $\|\textbf{x}\|^{2}:=\sum_{i}x_{i}^{2}$ for the Bloch vector $\textbf{x}=(x_{1},x_{2},x_{3})^{t}$. $\lambda_{max}$  is the maximum eigenvalue of
$x x^{t}+T T^{t}$. The elements of the correlation matrix and of the Bloch vector can be obtained from the following relations,
\begin{equation}\label{matrix}
\begin{split}
t_{ij}&=\frac{1}{2}\text{tr}(\rho_{AB}\sigma_{i}\otimes{\sigma_{j}}), \quad i,j=1,2,3\\
x_{i}&=\frac{1}{2}\text{tr}(\rho_{AB}\sigma_{i}\otimes I),\quad i=1,2,3
\end{split}
\end{equation}
Since both the geometric discord and MIN attains same highest value $0.5$ for Bell states, we did not use any normalization factor in their respective definitions.

\section{Evolution of MIN and Geometric Discord under markovian dynamics}
Generally, a quantum channel maps a quantum state to another quantum state. Quantum channel is actually a trace preserving, completely positive map. For example, unitary time evolution of a closed quantum system is a quantum map. Any completely positive quantum map admits a Kraus decomposition. An quantum channel is called unitary if it maps completely mixed state to a completely mixed state. In particular, unitary channels are unital and due to quantum analogue of Birkhoff's theorem, for qubit system any unital channel can be written as convex combination of unitary channels.  Let us consider a generic qubit density matrix $\rho=\frac{1}{2}(\mathbb{I}+\,\overrightarrow{v}.\overrightarrow{\sigma})$ with Bloch vector $\overrightarrow{v}\in \mathbb{R}^3$ such that $|\overrightarrow{v}|\leq 1$. Under the action of a general quantum channel $\Phi$, the state will evolve to $\Phi(\rho)=\frac{1}{2}(\mathbb{I}+\,\overrightarrow{v'}.\overrightarrow{\sigma})$ where $\overrightarrow{v}\rightarrow\overrightarrow{v'}=T\overrightarrow{v}+\overrightarrow{t}$ with $T$ a real $3\times 3$ matrix. For qubit unital channel $\overrightarrow{t}=\mathbf{0}$. Hence, unital channel, acting on a qubit, has a simple one-to-one parametrization in terms of a real matrix $T$. An quantum channel is said to be Markovian if it is a solution of a master equation with Lindblad type generator. We will consider here some important unital and non-unital channels in the context of our work.

Now let us consider two qubits, interacting independently with the individual environments. Then their evolution can be described by the Lindblad equation. This can be written in terms of Kraus operators in the form
\begin{equation}
\rho_{AB}(t)=\sum_{\mu,\nu}E_{\mu,\nu}\rho_{AB}(0)E_{\mu,\nu}^{\dag}
\end{equation}
where $\rho_{AB}(0)$ is the initial state shared between parties A and B, $\rho_{AB}(t)$ denotes the state after time $t$ and $E_{\mu,\nu}$ are the Kraus operators which satisfies the trace preserving relation $\sum_{\mu,\nu}E_{\mu,\nu}^{\dag}E_{\mu,\nu}=I$ for all $t$.
Let us consider the initial state as the X state of the form
\[ \rho_{AB}(0)=\left[ \begin{array}{cccc}
 \rho_{11}          & 0             & 0          & \rho_{14}    \\
  0                 & \rho_{22}     & \rho_{23}  & 0            \\
  0                 & \rho_{23} & \rho_{33}  & 0            \\
  \rho_{14}    & 0             & 0          & \rho_{44}    \\
\end{array} \right]\]
with the conditions $\sum \rho_{ii}=1$ and proper positivity constraints. We have taken each entries as non-negative because  by local unitary transformation the elements can be made real positive without affecting the discord or MIN since both are invariant under local unitary transformation. Hence we need ultimately $5$ parameters to describe any state from this class. For studying the evolution of correlations, here we will consider three kinds of initial states and all of them belong to this class of states and they all can be described by a single parameter.
\begin{eqnarray}
  \rho_{1}&=&|\psi\rangle\langle\psi|\:\text{where}\, |\psi\rangle =\sqrt{1-\alpha}|00\rangle+\sqrt{\alpha}|11\rangle,\alpha\in[0,1]\label{first} \\
  \rho_{2}&=&\frac{1-\alpha}{4}I+\alpha|\psi^{-}\rangle\langle\psi^{-}|, \alpha\in[0,1]\label{second} \\
  \rho_{3}&=&\frac{\alpha}{2}(|00\rangle\langle00|+|11\rangle\langle11|)+(1-\alpha)|\psi^{-}\rangle\langle\psi^{-}|,\alpha\in[0,1]\label{third}
\end{eqnarray}
The first one is a pure state, second one is a Werner state and third one is generalized Vedral-Plenio state which is a mixture of a  mixed state with the singlet state ($|\psi^{-}\rangle=\frac{1}{\sqrt{2}}(|01\rangle-|10\rangle))$. All these three kind of states are basically X states. The important factor here is the X state structure of all these three states and remain unchanged under the action of the decoherence channels. We will consider that both the qubits decay at the same rate under each decoherence channel.\\

\textbf{Depolarizing Channel}: Application of this type of channel depolarize the density matrix to completely mixed state with probability $p$. In this case, the Kraus operators have the form: $E_{0}=\sqrt{1-\frac{3 \gamma}{4}}I, E_{1}=\sqrt{\frac{\gamma}{2}}X, E_{2}=\sqrt{\frac{\gamma}{2}}Y, E_{3}=\sqrt{\frac{\gamma}{2}}Z$ with $(X, Y, Z)\equiv(\sigma_{x},i\sigma_{y},\sigma_{z})$.
Consider the Initial condition be $\rho_{AB}(0)=\rho_{1} (\ref{first})$.
Under the independent action of the channel on both the qubits the density matrix elements evolve as,
\begin{eqnarray}
\begin{split}
\rho_{11}(t)&=\rho_{11}(0)(1-\gamma)+\frac{\gamma^{2}}{4}\\
\rho_{22}(t)&=\rho_{33}(t)=\frac{\gamma}{2}(1-\frac{\gamma}{2})\\
\rho_{44}(t)&=1-\rho_{11}(t)-2\rho_{22}(t)\\
\rho_{14}(t)&=\rho_{41}(t)=\rho_{14}(0) (1-\gamma)^{2}\\
\end{split}
\end{eqnarray}
The evolution of the Bloch vector and the correlation matrix can be obtained from the relations in (\ref{matrix}). MIN and Geometric discord for this initial state is obtained as
\begin{equation}
N(\rho_{AB}(t))=D_{g}(\rho_{AB}(t))=2\alpha(1-\alpha)(1-\gamma)^{4}
\end{equation}
If the initial state be of the form (\ref{second}) then noting that $\|x(t)\|=0 \, ~\forall t $,  we have
\begin{equation}
D_{g}(\rho_{AB}(t))=N(\rho_{AB}(t))=\frac{\alpha^{2}}{2}(1-\gamma)^{4}
\end{equation}
\begin{figure}[!htb]
\scalebox{0.4}
{\includegraphics{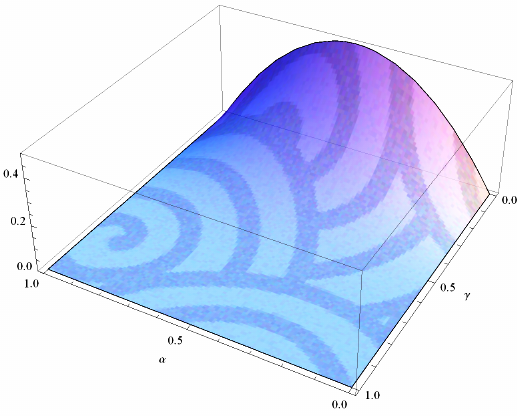}}
\label{Fig:I}
\caption{\textit{(Color Online) Monotonic nature of decay of MIN for the pure qubit initial state under independent depolarising noise. For $\alpha=0$ or $1$, MIN remains zero always.}}
\end{figure}
If we consider $\gamma=1-\exp[-\Gamma_{1} t]$, i.e., $\Gamma_{1}$ is the rate of depolarizing then we observe the monotonic nature of decay of MIN and geometric discord as $\gamma\rightarrow 1(\equiv t\rightarrow \infty)$.\\

\textbf{Dephasing Channel}: Physically, dephasing/phase damping corresponds to any process of losing coherence without any exchange of energy. Kraus operators corresponding to this process are given by
\[ E_{0}=\left[ \begin{array}{cc}
 1 & 0 \\
  0 & \sqrt{1-\gamma} \\
\end{array} \right], E_{1}=\left[ \begin{array}{cc}
 0 & 0 \\
  0 & \sqrt{\gamma} \\
\end{array} \right]\]
Under the action of this channel the density matrix elements corresponding to the initial state $\rho_{AB}(0)=\rho_{1}$ (\ref{first}) evolve as
\begin{eqnarray}
\begin{split}
\rho_{ii}(t)&=\rho_{ii}(0)\quad \text{for all}\quad i=1,2,3,4\\
\rho_{14}(t)&=(1-\gamma)\rho_{14}(0) \\
\end{split}
\end{eqnarray}
MIN and discord for this state is given by
\begin{equation}\label{dephasing1}
\begin{split}
N(\rho_{AB}(t))&=
\begin{cases}
\frac{1}{4}+\frac{(1-\gamma)^2}{4}&  \text{if}~\alpha= \frac{1}{2}\\
2\alpha (1-\alpha)(1-\gamma)^2& \text{if}~\alpha\neq\frac{1}{2}\\
\end{cases}\\
D_{g}(\rho_{AB}(t))&=2\alpha (1-\alpha)(1-\gamma)^2
\end{split}
\end{equation}
we take $\gamma=1-\exp[-\Gamma_{2} t]$ where $\Gamma_{2}$ is the rate of dephasing.
\begin{figure}[!htb]
\scalebox{0.5}
{\includegraphics{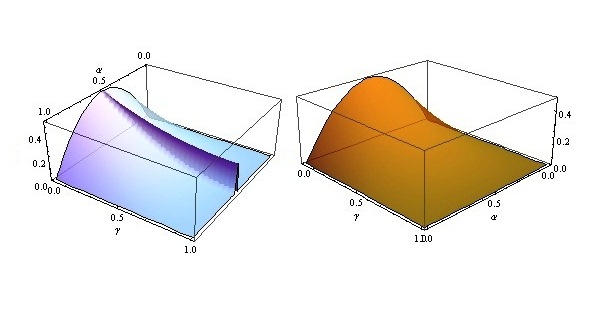}}
\label{Fig:II}
\caption{\textit{(Color Online) The nature of decay of MIN and geometric discord and respectively for the pure initial state under independent dephasing noise. While the value of discord decays monotonically for all $\alpha$, MIN for $\alpha=0.5$ does not vanishes even in infinite time.}}
\end{figure}

Again, if the initial condition be of the form $\rho_{AB}(0)=\rho_{2}$ (\ref{second}) then,
\begin{eqnarray}
\begin{split}
\rho_{ii}(t)&=\rho_{ii}(0)\quad \text{for all}\quad i=1,2,3,4\\
\rho_{23}(t)&=\rho_{23}(0) (1-\gamma)\\
\end{split}
\end{eqnarray}
MIN and discord for this state is given by,
\begin{equation}\label{dephasing2}
\begin{split}
N(\rho_{AB}(t))&=(\frac{\alpha}{2})^{2}(1-\gamma)^{2}+(\frac{\alpha}{2})^{2}\\
D_{g}(\rho_{AB}(t))&=2(\frac{\alpha}{2})^{2}(1-\gamma)^2
\end{split}
\end{equation}

\begin{figure}[!htb]
\scalebox{0.5}
{\includegraphics{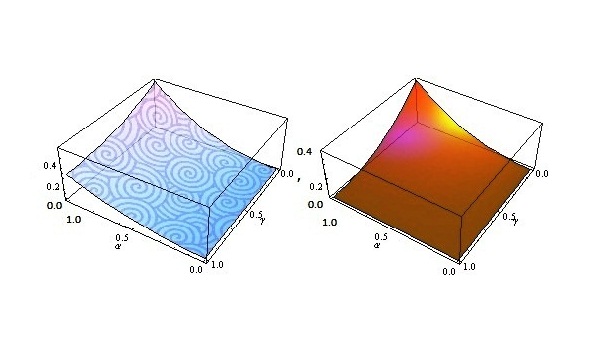}}
\label{Fig:III}\\
\caption{\textit{(Color Online) The nature of decay of MIN and geometric discord and respectively for  the Werner state under independent dephasing noise. MIN shows more robustness against dephasing noise than geometric discord.}}
\end{figure}

From FIG.2 we see that MIN for maximally entangled state does not vanishes as $\gamma\rightarrow 1$. So under dephasing noise while the quantumness vanishes, nonlocality remains present throughout the process for this state. In this sense nonlocality is more robust than quantumness. Similar kind of conclusion can be obtained for the Werner state (FIG.3). In this case for any non-zero $\alpha$, MIN does not decay to zero like geometric discord. Decay of MIN in case of Werner states clearly suggests that Werner state is better candidate than pure state for preserving MIN in any dephasing process.\\

\textbf{Generalized Amplitude Damping Channel}: Exchange of energy occurs in this type of decoherence noise. Kraus operators corresponding to this channel are given by,
\[ E_{0}=\sqrt{p}\left[ \begin{array}{cc}
 1 & 0 \\
  0 & \sqrt{1-\gamma} \\
\end{array} \right], E_{1}=\sqrt{p}\left[ \begin{array}{cc}
0 & \sqrt{\gamma} \\
  0 & 0 \\
\end{array} \right]\]

\[ E_{2}=\sqrt{1-p}\left[ \begin{array}{cc}
 \sqrt{1-\gamma} & 0 \\
  0 & 0 \\
\end{array} \right], E_{3}=\sqrt{1-p}\left[ \begin{array}{cc}
 0 & 0 \\
  \sqrt{\gamma} & 0 \\
\end{array} \right]\]

Evolution of the matrix element for the initial state $\rho_{AB}(0)=\rho_{1} $ (\ref{first}) can be written as,
\begin{equation}
\begin{split}
\rho_{11}(t)&=(1-(1-p)\gamma)^{2}-\alpha(1-\gamma)(1-(1-2p)\gamma)\\
\rho_{22}(t)&=-\gamma(-1+\alpha+p(1-2\alpha(1-\gamma)-2\gamma)+\gamma(1+p^{2}-\alpha))\\
\rho_{33}(t)&=\rho_{22}(t)\\
\rho_{44}(t)&=1-\rho_{11}(t)-2\rho_{22}(t)\\
\rho_{14}(t)&=\sqrt{\alpha-\alpha^{2}}(1-\gamma)\\
\end{split}
\end{equation}
and the dynamics of the density matrix elements for the initial condition $\rho_{AB}(0)=\rho_{2}$ (\ref{second}) is given by,
\begin{equation}
\begin{split}
\rho_{11}(t)&=\frac{1}{4}(-\alpha(1-\gamma)^{2}+(1+(2p-1)\gamma)^{2})\\
\rho_{22}(t)&=\rho_{33}(t)=\frac{1}{4}(1+\alpha(1-\gamma)^{2}-(1-2p)^{2}\gamma)^{2})\\
\rho_{44}(t)&=1-\rho_{11}(t)-2\rho_{22}(t)\\
\rho_{23}(t)&=\frac{\alpha}{2}(1-\gamma)\\
\end{split}
\end{equation}

We have plotted the figures corresponding to the initial states $\rho_{1} $ (\ref{first}) and $\rho_{2}$ (\ref{second}) considering three values of $p$ and $\gamma=1-\exp[-\Gamma_{3}t]$. $\Gamma_{3}$ is the rate of damping. $p=1$ corresponds to the usual amplitude damping channel. We have also compared the dynamics with the corresponding dynamics of geometric discord in the figures (FIG.4, FIG.5). From the figures, we observe monotonic nature of decay of MIN like geometric discord. No sudden death type behavior for MIN occurs. Clearly  MIN does not disappear in finite time. While the decay in the case of MIN is smooth, the corresponding decay in case of discord is not so smooth. There are certain regions where we can notice changing behavior of discord. Another important observation is that application of the amplitude damping channel completely decays the correlations as $\gamma\rightarrow 1$ for both the initial states. In case of pure state, when $\alpha=0$ or $1$, MIN remains zero always.\\

\begin{figure}[!htb]
\scalebox{0.5}
{\includegraphics{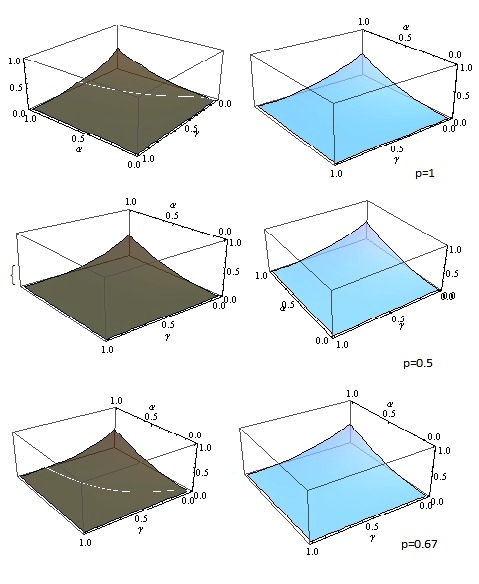}}
\label{Fig:IV}
\caption{\textit{(Color Online) Monotonic nature of decay of geometric discord and MIN respectively for the Werner state under independent generalized amplitude damping for  $p=1,0.5,0.67$ respectively. The white curve in the figures of discord shows its changing nature while the decay of MIN remains smooth.}}
\end{figure}

\begin{figure}[!htb]
\scalebox{0.5}
{\includegraphics{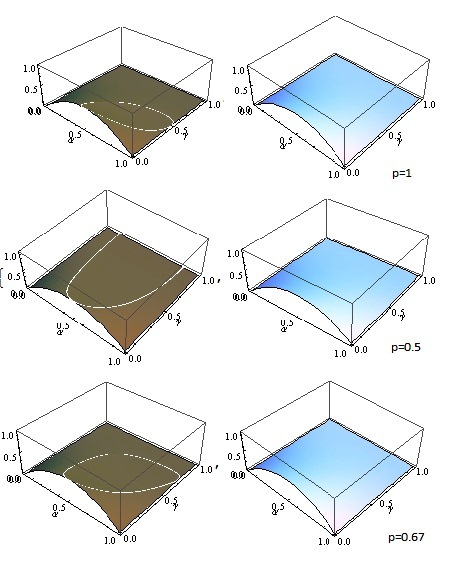}}
\label{Fig:V}
\caption{ \textit{(Color Online) Monotonic nature of decay of geometric discord and MIN respectively for the pure state under independent generalized amplitude damping  for $p=1,0.5,0.67$ respectively. Discord shows its changing nature while MIN remains smooth.}}
\end{figure}

\textbf{Dephasing and Amplitude damping}: Now let us consider the scenario where both the qubits are acted simultaneously by both the dephasing and amplitude damping noise. Since the diagonal elements remains unchanged under dephasing noise, so, in this case the off diagonal elements decay as the sum of the individual rate of decay. Here we consider that both the channel decays at the same rate, $\Gamma_{2}=\Gamma_{3}=\Gamma$. Under this consideration, the dynamics of density matrix elements for the initial state $\rho_{AB}(0)=\rho_{1}$ (\ref{first}) is,
\begin{equation}
\begin{split}
\rho_{11}(t)&=1-\alpha+\alpha(1-\exp[-\Gamma t])^{2}\\
\rho_{22}(t)&=\alpha(1-\exp[-\Gamma t])^{2}\exp[-\Gamma t]\\
\rho_{33}(t)&=\rho_{22}(t)\\
\rho_{44}(t)&=\alpha \exp[-2\Gamma t]\\
\rho_{23}(t)&=\sqrt{(\alpha-\alpha^{2})}\exp[-2\Gamma t]\\
\end{split}
\end{equation}\\

\begin{figure}[!htb]
\scalebox{0.5}
{\includegraphics{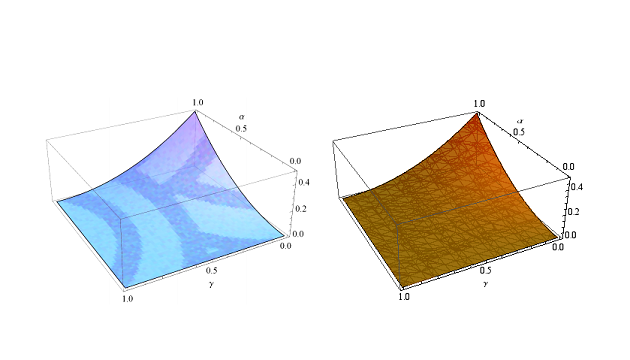}}
\label{Fig:VI}
\caption{\textit{(Color Online)Monotonic nature of decay geometric discord and MIN respectively for pure qubit state assuming simultaneous action of both the dephasing and amplitude damping channels ($p=1$) in each qubit.}}
\end{figure}

In FIG.6, we have plotted the values of MIN and compared with geometric discord. The nature of decay reveals additivity in the nature of decay of MIN under simultaneous action of both the noises. Geometric discord also shows the same behavior. We also compare the rate of decay of MIN for all the three types of decoherence channels, described earlier, for the Bell state. From FIG.8 we observe that MIN is more robust under dephasing noise than other two types of noise. In fact dephasing noise can not fully destroy this correlation in infinite run.

\begin{figure}[!htb]
\centering
\scalebox{0.35}
{\includegraphics{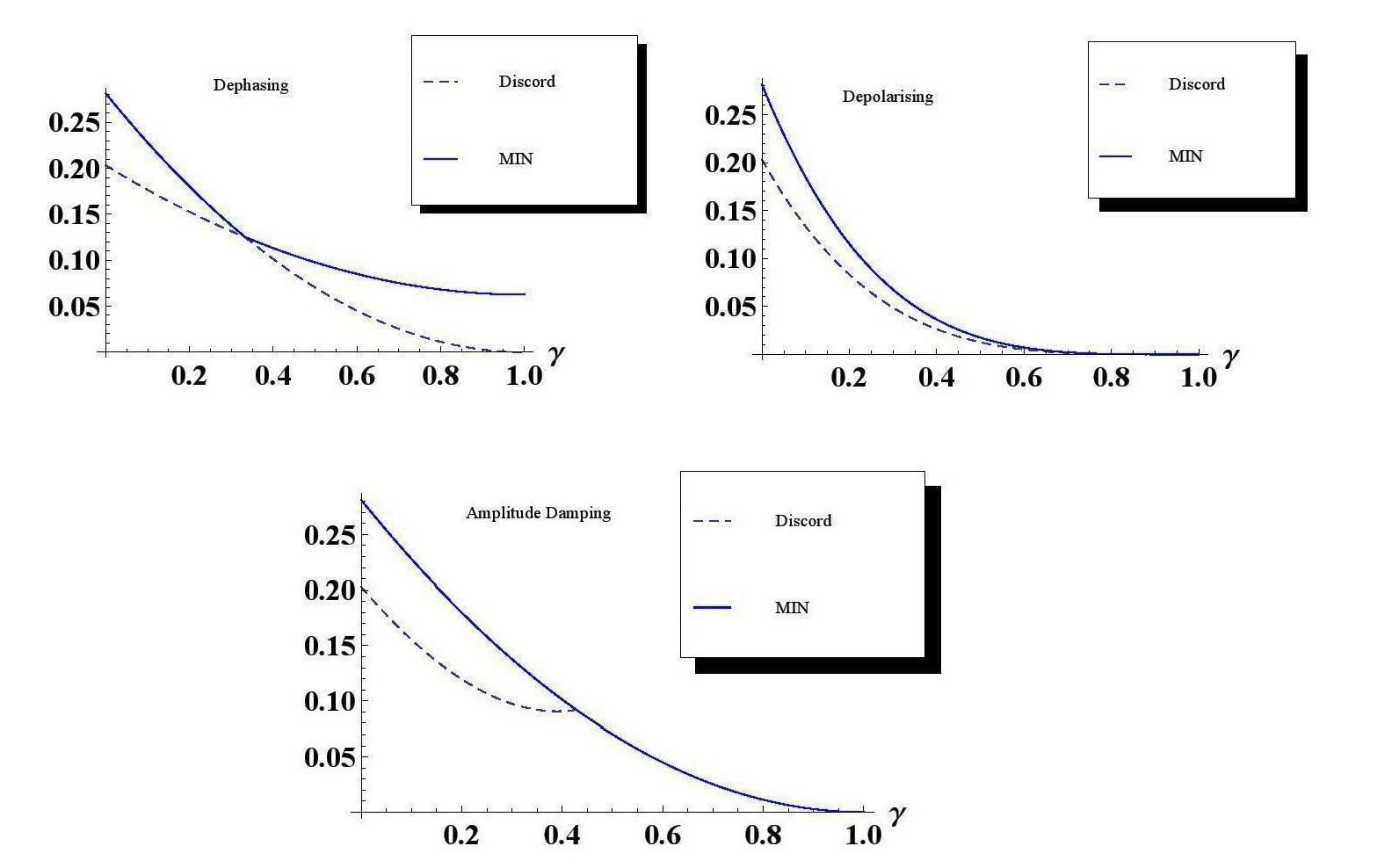}}
\caption{\textit{(Color Online)Comparison of the rate of decay of MIN and geometric discord under three types of decoherence channels for the initial state $\rho_{3}$ at $\alpha=0.25$}}
\label{Fig:VII}
\end{figure}

So far we have discussed dissipative dynamics for the initial states $\rho_{1}$ (\ref{first}) and $\rho_{2}$  (\ref{second}). Similar type of analysis can also be done by taking the initial state $\rho_{3}$ (\ref{third}). In the FIG.7, we have plotted the decay nature corresponding to this initial state $\rho_{3}$ for a fixed $\alpha$. From the figure, we observe that the dephasing noise can not decay MIN to zero in this case also. But there is a particular time when discord and MIN matches exactly and after that discord falls faster and ultimately goes to zero. Under dephasing noise discord always remains below than MIN and they both decays to zero. In case of amplitude damping noise discord falls slowly than MIN and at one time their values become same and ultimately they both decay at the same rate. We have compared the rate of decay of MIN for Bell state under the three channels in FIG. 8.\\
\begin{figure}[!htb]
\centering
\scalebox{0.5}
{\includegraphics{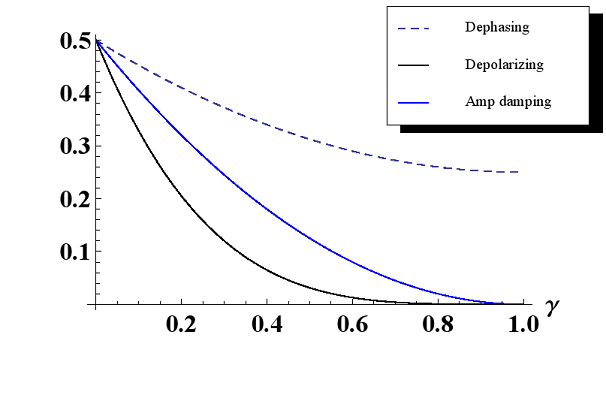}}
\caption{\textit{(Color Online) Comparison of the rate of decay of MIN for the Bell state $|\phi^{+}\rangle$ under the three decoherence channels.}}
 \label{Fig:VIII}
\end{figure}

\section{Dephasing and Amplitude damping noise under non-markovian dynamics}
Let us consider two non interacting qubits which are interacting independently with their environment. The interaction for each qubit with the environment for amplitude and dephasing noise can be modeled respectively by the Hamiltonian,
\begin{equation}\label{ham1}
\begin{split}
    H_{I}^{amplitude}=\sum_{k}(g_{k}\sigma_{-}b_{k}^{\dag}+g_{k}^{*}\sigma_{+}b_{k})\\
\end{split}
\end{equation}
\begin{equation}\label{ham2}
    H_{I}^{dephasing}=\sum_{k}\sigma_{z}(g_{k}b_{k}^{\dag}+g_{k}^{*}b_{k})\\
\end{equation}
where $g_{k}$'s are the coupling constants and $\sigma_{\pm}$ are the raising/lowering operators for qubit levels. The total Hamiltonian for each qubit reads (as in \cite{wu}),
\begin{equation}\label{totham}
    H^{total}=\frac{\omega_{0}}{2}\sigma_{z}+\sum_{k} \omega_{k}b_{k}^{\dag}b_{k}+H_{I}
\end{equation}
Here $\omega_{0}$ is the energy separation between the qubit level $|\pm\rangle$ and $b_{k}$ is the annihilation operator for the oscillator mode with frequency $\omega_{k}$. The operation elements for the dephasing noise in the basis $\{|+\rangle,|-\rangle\}$ are,
\[ E_{0}^{d}=\left[ \begin{array}{cc}
 p_{d}(t) & 0 \\
  0 & 1 \\
\end{array} \right], E_{1}^{d}=\left[ \begin{array}{cc}
 q_{d}(t) & 0 \\
  0 & 0 \\
\end{array} \right]\]
where $q_{d}(t)=\sqrt{1-|p_{d}(t)|^{2}}$ and $p_{d}(t)$ is the solution of the equation $\frac{d}{dt}p_{d}(t)=-\int_{0}^{t}d\tau f(t-\tau)p{d}(\tau)$ with $f$ as the noise correlation function as in \cite{wu}.
Similar operator elements for the amplitude damping noise in the same basis are,
\[ E_{0}^{a}=\left[ \begin{array}{cc}
 p_{a}(t) & 0 \\
  0 & 1 \\
\end{array} \right], E_{1}^{a}=\left[ \begin{array}{cc}
 0 & 0 \\
  q_{a}(t) & 0 \\
\end{array} \right]\]
Here $q_{a}(t)=\sqrt{1-p_{a}(t)^{2}}$, $p_{a}(t)=\exp\{\Gamma(t)\}$ and $\Gamma(t)$ is a real positive function of time as in \cite{wu}. We will consider Lorentzian spectral function $J(\omega)$ for the coupling and it has the form,
\begin{equation}
J(\omega)=\frac{1}{2\pi}\frac{\gamma \lambda^{2}}{(\omega-\omega_{0}+\Delta^{2})^{2}+\lambda^{2}}
\end{equation}
$\gamma$ is the decay rate of the upper qubit level $|+\rangle$, $\lambda$ is the coupling bandwidth and $\Delta$ is the detuning from the resonance frequency $\omega_{0}$.
Now consider a general two qubit X state in $\{|+\rangle,|-\rangle\}$ basis,
\[ \rho_{AB}(0)=\left[ \begin{array}{cccc}
 \rho_{11}          & 0             & 0          & \rho_{14}    \\
  0                 & \rho_{22}     & \rho_{23}  & 0            \\
  0                 & \rho_{23} & \rho_{33}  & 0            \\
  \rho_{14}    & 0             & 0          & \rho_{44}    \\
\end{array} \right]\]
with the conditions $\sum \rho_{ii}=1 $ and proper positivity constraints. We choose this basis  $\{|+\rangle,|-\rangle\}$ since the X state structure of the state $\rho(0)$ remains preserved under both the channel. The evolution of the density matrix elements under these two types of noises can be seen as\\

\textbf{Amplitude Damping:}
\begin{eqnarray}
\begin{split}
\rho_{11}(t)&=|p_{a}(t)|^{4}\rho_{11}(0)\\
\rho_{22}(t)&=|p_{a}(t)|^{2}(\rho_{22}(0)+ q_{a}^{2}(t)\rho_{33}(0))\\
\rho_{33}(t)&=|p_{a}(t)|^{2}(\rho_{33}(0)+ q_{a}^{2}(t)\rho_{11}(0))\\
\rho_{44}(t)&=1-\rho_{11}(0)-\rho_{22}(0)-\rho_{33}(0)\\
\rho_{14}(t)&=|p_{a}(t)|^{2}\rho_{14}(0)\\
\rho_{23}(t)&=|p_{a}(t)|^{2}\rho_{23}(0)\\
\end{split}
\end{eqnarray}

\textbf{Dephasing Noise:}
\begin{eqnarray}
\begin{split}
\rho_{ii}(t)&=\rho_{ii}(0),i=1,2,3,4\\
\rho_{14}(t)&=p_{d}^{2}(t)\rho_{14}(0)\\
\rho_{23}(t)&=p^{2}_{d}(t)\rho_{23}(0)\\
\end{split}
\end{eqnarray}\\
\begin{figure}[!htb]

\begin{minipage}[b]{0.45\linewidth}
\centering
  \includegraphics[width=.4\linewidth]{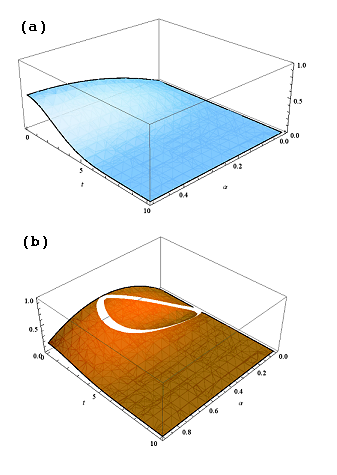}
  \caption{A \textsl{(Color Online) Evolution of (a) MIN and (b)geometric discord respectively for amplitude damping noise for the pure initial state $\sqrt{1-\alpha}|00\rangle+\sqrt{\alpha}|11\rangle$. Here we take Lorentzian spectral function (ref \cite{wu}) for qubit environment coupling. We plot both the MIN and geometric discord for a fixed coupling bandwidth $\lambda=0.1\gamma$ and fixed central frequency $\omega=\gamma$. The time axis is in the units of $1/\gamma$}}
  \label{fig:sub1}
\end{minipage}
\hspace{0.5cm}

\begin{minipage}[b]{0.45\linewidth}
  \centering
  \includegraphics[width=.4\linewidth]{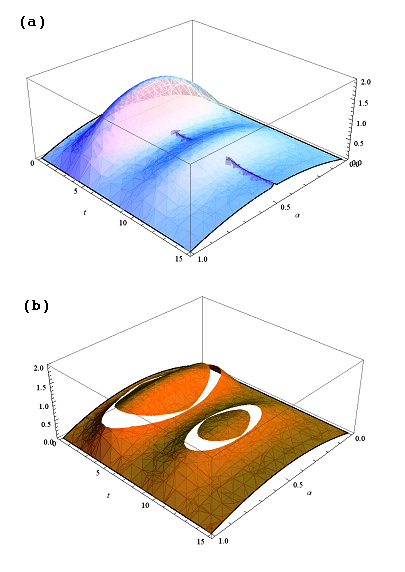}
  \caption{(Color Online) \textsl{Evolution of (a) MIN and (b) geometric discord respectively for phase damping noise for pure initial state $\sqrt{1-\alpha}|00\rangle+\sqrt{\alpha}|11\rangle$. Here we take Lorentzian spectral function (ref \cite{wu}) for qubit environment coupling as previous case. We plot both the MIN and geometric discord for a fixed coupling bandwidth $\lambda=0.1\gamma$ and fixed detuning $\Delta=0.01\gamma$. The time axis is in the units of $1/\gamma$}}
  \label{fig:sub2}
\end{minipage}
\end{figure}

In both the cases we have taken the identical noise on each qubit, i.e., $p_{d/a}^{\text{A}}=p_{d/a}^{\text{B}}=p_{d/a}$. MIN and geometric discord of both the evolved states can be obtained by the relations (\ref{min}),(\ref{discord}). But due to their cumbersome form we concentrate only on a few classes as mentioned earlier in Markovian part. For the phase damping case we take pure initial state as of the form (\ref{first}). MIN and geometric discord of the evolved state can be obtained easily and found to be of the following form,

\begin{equation}\label{minnonmarkovphase}
N(\rho_{AB})=
\begin{cases}
2|p_{d}(t)|^{4}(\alpha-\alpha^{2}) & \text{if}~\alpha \neq 0.5\\
0.25+2|p_{d}(t)|^{4}(\alpha-\alpha^{2})-\min\{0.25,|p_{d}(t)|^{4}(\alpha-\alpha^{2})\}& \text{if}~\alpha=0.5\\
\end{cases}
\end{equation}
\begin{equation}\label{disnonmarkovphase}
\begin{split}
D(\rho_{AB})=2|p_{d}(t)|^{4}(\alpha-\alpha^{2})+0.25+(0.5-\alpha)^{2}-\max\{0.5-\alpha+\alpha^{2},|p_{d}(t)|^{4}(\alpha-\alpha^{2})\}
\end{split}
\end{equation}

Similar quantities under amplitude damping case can be written for initial pure state evolution. However we skip them from writing due to their clumsy form. Instead we plotted them in Figure (9) and (10) and compared their form of decay.\\
\section{Conclusion}
Thus we have compared two different types of correlations of two qubit quantum systems under Markovian and non-Markovian decoherence dynamics. In the Markovian case, we have found additive dynamics of MIN under simultaneous action of decoherence noise. Geometric discord also behaves similarly. We have also shown some states which retains their nonlocality even after applying a type of decoherence noise. Such behavior is not observed for geometric discord for those states. Under amplitude damping noise Sudden change in the decay of geometric discord is also observed but in this case MIN decays smoothly without any change. Lastly, we have analyzed the effect of non-Markovian dynamics on MIN and geometric discord for a initial pure state and checked their forms of decay graphically. \\

{\bf Acknowledgement.} The author A. Sen acknowledges the financial support from University Grants Commission, New Delhi, India.


\begin{thebibliography}{1}
\bibitem{zurek}H. Ollivier and W. H. Zurek, Phys. Rev. Lett. \textbf{88}, 017901 (2001).
\bibitem{dakic}B. Dakic, V. Vedral and C. Brukner,Phys. Rev. Lett.  \textbf{105}, 190502(2010).
\bibitem{luo} S. Luo and S. S. Fu, Phys. Rev. Lett. \textbf{106}, 120401 (2011).
\bibitem{guo}Y. Guo, and J.Hou, arXiv:1107.0355v1.
\bibitem{eberly} T. Yu and J.H. Eberly, Phys. Rev. Lett.  \textbf{97}, 140403 (2006).
\bibitem{werlang}T. Werlang et.al., Phys. Rev. A \textbf{80}, 024103 (2000).
\bibitem{ferraro}A. Ferraro, L. Aolita, D. Cavalcanti, F. M. Cucchietti, A. Acin, Phys. Rev. A \textbf{81},052318 (2009).
\bibitem{maziero} J. Maziero, L. C. Celeri, R. M. Serra, V. Vedral, Phys. Rev. A \textbf{80},044102 (2009).
\bibitem{bellomo}B. Bellomo, R. L. Franco, G. Compagno, Phys. Rev. A \textbf{86}, 012312 (2012).
\bibitem{2} A. Streltsov, G. Adesso, M. Piani and D. Bruss, Phys. Rev. Lett. \textbf{109}, 050503 (2012).
\bibitem{Guo-Feng} Guo-Feng Zhang, Heng Fan, Ai-Ling Ji, Wu-Ming Liu, arXiv:1201.1949.
\bibitem{Ming} Ming-Liang Hu and Heng Fan,arXiv:1201.6430v2
\bibitem{1} R. Auccaise, L. C. Céleri, D. O. Soares-Pinto, E. R. deAzevedo, J. Maziero, A. M. Souza, T. J. Bonagamba, R. S. Sarthour, I. S. Oliveira, and R. M. Serra, Phys. Rev. Lett. \textbf{107}, 140403 (2011).
\bibitem{3} J.L. Guo, Y.J. Mi and H.S. Song, Eur. Phys. J. D \textbf{66}, 24 (2012).
\bibitem{4} M. Ramzan, Quantum Information Processing, (2013), doi: 10.1007/s11128-013-0558-0.
\bibitem{5} S. Alipour, A. Mani, A. T. Rezakhani, Phys. Rev. A \textbf{85}, 052108 (2012).
\bibitem{6} Rosario Lo Franco, Bruno Bellomo, Sabrina Maniscalco, Giuseppe Compagno, Int. J. Mod. Phys. B \textbf{27}, 1245053 (2013).
\bibitem{mirafzali} S. Y. Mirafzali, I. Sargolzahi, A. Ahanj, K. Javidan, and M. Sarbishaei, arXiv:1110.3499.
\bibitem{xi} Z. Xi, X. Wang and Y. Li, arXiv:1112.0736v1.
\bibitem{wu} Shin-Tza Wu, Chinese Journal of Physics, VOL. \textbf{50}, NO. 1 (2012)
\end{thebibliography}
\end{document}